\documentclass[prb,twocolumn,aps,showpacs,fixfloats]{revtex4}
\usepackage{graphicx}
\usepackage{bm}
\usepackage{amsmath,amssymb}
\usepackage{subfigure}
\usepackage{float}
\usepackage{latexsym}
\usepackage{color}
\usepackage{enumerate}
\usepackage{pdfpages}
\usepackage{tikz}
\usepackage{hyperref}
\usepackage{graphicx}
\usepackage{bm}
\usepackage{amssymb} 
\usepackage{amsmath}
\usepackage{subfigure}
\usepackage{relsize}

\begin{document}


\title{ Size quantization of an exciton: A toy model of the ``dead layer" }

\author{M. E. Raikh}

\affiliation{ Department of Physics and
Astronomy, University of Utah, Salt Lake City, UT 84112}

\begin{abstract}
Size-quantization levels of an exciton in large nanocrystals
 is studied theoretically. For the nanocrystal size, $L$, much bigger
 than the Bohr radius, $a_B$, the level positions do not depend on $a_B$.
 The correction to the levels in a small parameter $a_B/L$ depends on the
reflection phase of the exciton from the boundary. Calculation of this phase
constitutes a three-body problem: electron, hole, and the boundary. This calculation
can be performed analytically in the limit when the hole is much heavier than the
electron. Physically, a slow  motion of the hole towards the boundary takes place in the
effective potential created by the fast motion of the electron orbiting the hole
and touching the boundary.  As a result, the hole is reflected before reaching the boundary.
The distance of the closest approach of the hole to the boundary (the dead layer) exceeds $a_B$ {\em parametrically}.
\end{abstract}

\maketitle

\section{Introduction}

A concept of the exciton dead layer\cite{pekar,Hopfield} has emerged
in the course of study of the light reflection from a boundary between
air and a medium with strong exciton-photon coupling. As a result of
this coupling, two (exciton-like and photon-like) waves can propagate
in the medium. Thus, in order to find the reflection coefficient, conventional
boundary conditions of the continuity of the tangent components of electric
and magnetic fields, additional boundary condition is needed. Analysis in
the pioneering paper by Hopfield and Thomas\cite{Hopfield} suggests that
the form of this condition is vanishing of the exciton polarization at certain
distance, $l$, away from the boundary. This distance was estimated
in Ref. \onlinecite{Hopfield} as $l=2a_B$, where $a_B$ is the Bohr radius
of the exciton. In early, see e.g.
Refs. \onlinecite{1974,Bassani,DelSole2,DelSole1,DelSole,Bastard},
as well as in recent, see e.g. Refs. \onlinecite{Cardona,Ivchenko,Poddubny},
follow-up papers the concept of dead layer associated with distance, $l$,
was employed.

In the absence of the exciton-photon coupling,
the issue of reflection of the exciton from the surface
is still important. Namely, the {\it phase} of the reflection
coefficient defines the positions of the size-quantization
level of  the exciton in the quantum well.

In particular,
in Ref. \onlinecite{Bassani}, photoluminescence spectra
from thick GaAs quantum well revealed a number of
peaks attributed to the size quantization of excitons.
Their positions were fit by a formula
\begin{equation}
\label{formula}
{\cal E}_n=E_x+\frac{\pi^2\hbar^2n^2}{2M\bigl(L-2l\bigr)^2},
\end{equation}
where $L$ is the well thickness, $M$ is the net mass of the exciton, and $E_x$ is the position of the exciton line in the bulk GaAs. For the dead-layer thickness, the value
$l=a_B$ was chosen.

Size quantization also manifests itself in the positions of the exciton
luminescence lines in nanocrystals. These studies, pioneered in
Refs. \onlinecite{ekimov,Brus},  were later conducted on wide variety
of nanocrystals. The importance of dead layer for the size-quantization
in ZnO nanocrystals was studied in Ref. \onlinecite{Fonoberov}.

Recent revival of  interest to excitons in nanocrystals is
related to inorganic perovskites, which, due to their
exceptionally high photoemission rates, show a good promise
for applications. As demonstrated experimentally in
Ref. \onlinecite{nano1}, these nanocrystals
exhibit a strong size-quantization effect. Further
experiments, see e.g. Refs. \onlinecite{nano2,nano3,nano4,nano5},
indicated that nanocrystal sizes vary in the range $10$nm-$50$nm,
while the exciton Bohr radius was estimated as $a_B \sim 2-6$nm.\cite{nano5}

On the theoretical side, microscopic description of the exciton dead layer poses
a challenge, since it is a three-body problem: electron, hole and a boundary.
In other words, while the motion of electron and hole can be separated into
the motion of the center of mass and the relative motion, this separation
is inconsistent with the boundary conditions that the two-particle wave function
turns to zero when electron and hole touch the boundary {\em individually}.
These no-escape boundary conditions constitute a microscopic origin
of the dead layer.

Below we consider a toy model which allows to derive the dead layer analytically and confirm  Eq. (\ref{formula}) rigorously from the Schr{\"o}dinger equation. A crucial assumption,
which allows us to capture the electron-hole correlation in the presence of the boundary, is that the hole mass, $m_h$, is much bigger than the electron mass, $m_e$. Under this assumption, to the first
order in $m_e/m_h$, the fast motion of  electron takes place in the field of a ``static" hole.\cite{DelSole}
Our main conclusion is that, in the formula Eq.~(\ref{formula}), the thickness of the dead layer, $l_N$, is not constant, but depends on the number, $N$, of the quantization level.

\begin{figure}
\includegraphics[width=125mm]{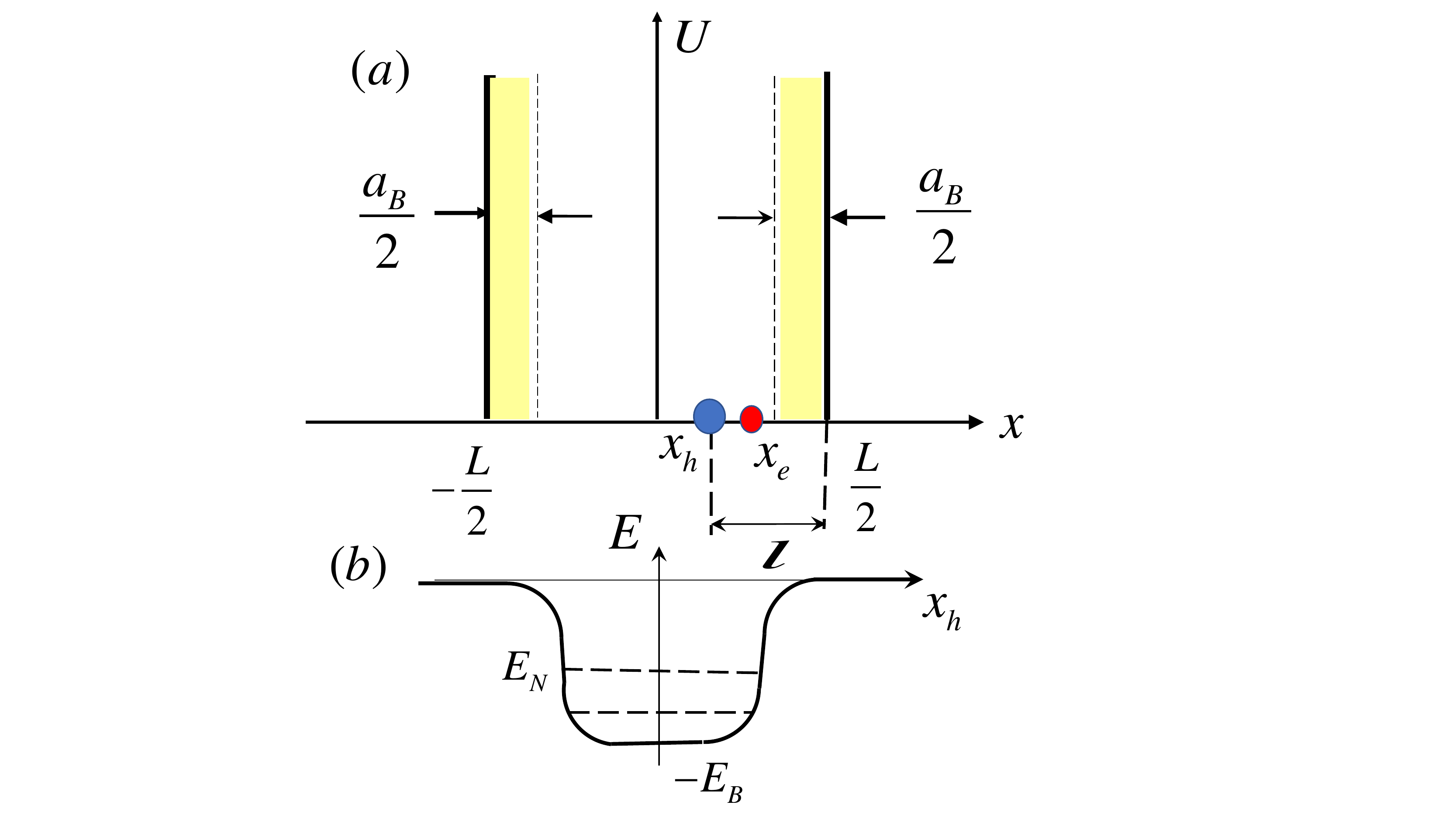}
\caption{(Color online) (a) Within a toy model, one-dimensional
nanocrystal occupies a domain $\vert x \vert <\frac{L}{2}$.
When the hole is much heavier than the electron, a slow  motion of the hole
 takes place in the effective potential created by the fast motion of
 the electron. As electron approaches one of the hard walls closer than $\frac{1}{2}a_B$,
  where $a_B$ is the Bohr radius, the bound state for electron disappears.
 When electron is outside the yellow regions, it ``senses"
the boundary only weakly, and, via the interaction, creates a barrier for the hole;
(b) Discrete energies of the exciton is determined by the size quantization of the hole.
The hole does not touch the boundaries, it is rather  reflected from the barrier of the
width, $l$, (the dead layer) created by the electron.}
\label{figure1}
\end{figure}

\section{The model}

To maximally simplify the calculations, we assume that the confinement is one-dimensional in the form of two hard walls at $x=\pm \frac{L}{2}$, see Fig. \ref{figure1}. Moreover, we will replace the Coulomb attraction between the electron and hole by a short-range attraction. Under these assumptions, the Hamiltonian of the model reads

\begin{equation}
\label{Hamiltonian}
{\hat H}=\frac {{\hat p}_{e}^2}{2m_e}+\frac {{\hat p}_{h}^2}{2m_h}-v_0\delta(x_e-x_h),
\end{equation}
where $x_e$ and $x_h$ are the coordinates of the electron and hole, respectively;~$v_0$ is the attraction strength.
The wave function $\Psi(x_e,x_h)$ satisfies the boundary conditions
\begin{equation}
\label{BoundaryConditions}
\Psi\left(\pm \frac{L}{2},x_h\right)=0,~~~~
\Psi\left(x_e,\pm \frac{L}{2}\right)=0.
\end{equation}
If $m_e$ and $m_h$ were comparable, the conditions Eq. (\ref{BoundaryConditions}) would
strongly mix the motion of the center of mass and the relative motion near the boundaries. Thus, to
proceed further, we assume that the hole is much heavier than
the electron and exploit the smallness of the parameter $m_e/m_h$.

\subsection{Lowest order in $m_e/m_h$}


In the limit of large $m_h$, we can neglect the kinetic energy  of
the hole. After that $x_h$ enters the Hamiltonian
Eq. (\ref{Hamiltonian}) only as a parameter, which can
be viewed as fixed.
Then the general form of the eigenfunction, $\varphi_{x_h}(x_e)$, which turns to zero at $x_e=\pm \frac{L}{2}$ is the following
\begin{eqnarray}
\label{FixedHole}
\varphi_{x_h}(x_e)=\alpha\sinh\Big[\kappa_h\Big(\frac{L}{2}+x_e\Big)\Big],~~~-\frac{L}{2}<x_e<x_h
\nonumber\\
\varphi_{x_h}(x_e)=\beta\sinh\Big[\kappa_h\Big(x_e-\frac{L}{2}\Big)\Big],~~~x_h<x_e<\frac{L}{2}.
\end{eqnarray}
The corresponding energy is $E_e(\kappa_h)=-\frac{\hbar^2 \kappa_h^2}{2m_e}$.
The condition of continuity at $x_e=x_h$ reads
\begin{equation}
\label{continuity}
\alpha\sinh\Big[\kappa_h\Big(\frac{L}{2}+x_h\Big)\Big]
=\beta\sinh\Big[\kappa_h\Big(x_h-\frac{L}{2}\Big)\Big].
\end{equation}
The discontinuity of the derivative at $x_e=x_h$ yields the second condition
\begin{align}
\label{discontinuity}
\kappa_h\alpha &\cosh\Big[\kappa_h\Big(\frac{L}{2}+x_h\Big)\Big]-
\kappa_h\beta\cosh\Big[\kappa_h\Big(x_h-\frac{L}{2}\Big)\Big]\nonumber\\
&=-\frac{2m_ev_0}{\hbar^2}\alpha \sinh\Big[\kappa_h\Big(\frac{L}{2}-x_h\Big)\Big].
\end{align}
Combining Eqs. (\ref{continuity}) and (\ref{discontinuity}), we get
the equation for $\kappa_h$
\begin{equation}
\label{equation}
\sinh \kappa_hL=\frac{2m_ev_0}{\hbar^2\kappa_h}\sinh \Big[\kappa_h\Big(\frac{L}{2}-x_h\Big)\Big]
\sinh \Big[\kappa_h\Big(\frac{L}{2}+x_h\Big)\Big].
\end{equation}
This equation defines the dependence $\kappa_h(x_h)$, i.e. the dependence of the electron energy on the position of the hole.

The concept of a dead layer is meaningful only when the ``nanocrystal" is big, namely, $L$ is much bigger than the size of the exciton.
In the limit $\kappa_hL \gg 1$ and $\big(\frac{L}{2}-x_h\big) \ll L$, we can replace
 $\sinh \kappa_hL$ by $\frac{1}{2}\exp \big(\kappa_hL\big)$ and
 $\sinh \big[\kappa_h\big(\frac{L}{2}+x_h\big)\big]$ by
 $\frac{1}{2}\exp \big[\kappa_h\big(\frac{L}{2}+x_h\big)\big]$. Then
 Eq. (\ref{equation}) assumes the form
\begin{equation}
\label{limit}
\kappa_h=\frac{m_ev_0}{\hbar^2}\bigg\{1-\exp\Big[-2\kappa_h\Big(\frac{L}{2}-x_h \Big)  \Big]    \bigg\},
\end{equation}
where the exponent in the square brackets accounts for electron ``hitting" the right wall and
does not ``sense" the left wall.
When the exponent is small, Eq. (\ref{limit}) yields
\begin{equation}
\label{reproduce}
\kappa_h=\frac{m_ev_0}{\hbar^2}=\frac{1}{a_B},
\end{equation}
and, correspondingly, $E_B=\frac{\hbar^2}{2m_ea_B^2}$ for the binding energy.

As $x_h$ approaches  the right (or left)  wall, the bound state disappears at critical
$\frac{L}{2}-x_h=\frac{a_B}{2}$. Near the threshold, the solution of Eq. (\ref{limit})
behaves linearly with $x_h$
\begin{equation}
\label{linearly}
\kappa_h=\frac{\Big(\frac{L}{2}-x_h-\frac{a_B}{2}   \Big)}{\left(\frac{L}{2}-x_h   \right)^2}
\approx \frac{4}{a_B^2}\left[\frac{L}{2}-x_h-\frac{a_B}{2}   \right].
\end{equation}
The corresponding normalized wave functions have the form

\begin{equation}
\label{normalized}
\varphi_{x_h}(x_e)=\frac{\left( 2\kappa_h  \right)^{1/2}}{a_B  } \begin{cases}\bigl(L-2x_e\bigr),~ x_h<x_e<\frac{L}{2}, \\
a_B\exp\bigl[\kappa_h(x_e-x_h)\bigl],~ x_e<x_h.
 \end{cases}
\end{equation}
With $E_e(\kappa_h)$ depending on $x_h$ via Eq. (\ref{limit}), a
hole with a {\em finite} mass will move slowly in the effective potential created by a fast-moving electron. We will establish the form of this potential  in the next Section.

\section{Effective potential for a hole}

To incorporate the hole motion, we search for the solution of the Scr{\"o}dinger equation  ${\hat H}(x_e,x_h)\Psi=E\Psi$
in the form
\begin{equation}
\label{form}
\Psi(x_e,x_h)=\varphi_{x_h}(x_e)\Phi(x_h).
\end{equation}
Substitution of this form
into the Schr{\"o}dinger equation yields
\begin{align}
\label{substitution}
-&\frac{\hbar^2}{2m_h}\left[\varphi_{x_h}(x_e)\frac{\partial^2\Phi }{\partial x_h^2}+
2\frac{\partial \varphi_{x_h}(x_e)}{\partial x_h} \frac{\partial \Phi}{\partial x_h}
+\frac{\partial^2\varphi_{x_h}(x_e)}{\partial x_h^2} \Phi(x_h)\right]\nonumber\\
=&\Big[E-E_e(\kappa_h)    \Big]\varphi_{x_h}(x_e)\Phi(x_h).
\end{align}
In deriving Eq. (\ref{substitution}) we took into
account that $\varphi_{x_h}(x_e)$ satisfies the equation
\begin{equation}
\label{xe}
\left[\frac {{\hat p}_{e}^2}{2m_e}-v_0\delta(x_e-x_h)\right]\varphi_{x_h}(x_e)=
E_e(\kappa_h)\varphi_{x_h}(x_e).
\end{equation}

We assume that the electron wave function $\varphi_{x_h}(x_e)$ is normalized
$\int dx_e \Big(\varphi_{x_h}(x_e)\Big)^2=1$. To obtain a closed equation for the
hole wave function, $\Phi(x_h)$, we multiply Eq.~(\ref{substitution}) by $\varphi_{x_h}(x_e)$
and integrate over $x_e$. This yields

\begin{equation}
\label{effective1}
-\frac{\hbar^2}{2m_h}\left[\frac{\partial^2\Phi }{\partial x_h^2}+
2I(x_h)\frac{\partial\Phi }{\partial x_h} +J(x_h)\Phi   \right]=\Big[E-E_e(\kappa_h)
\Big]\Phi,
\end{equation}
where the functions $I(x_h)$ and $J(x_h)$ are defined as
\begin{align}
\label{functions}
I(x_h)=&\int dx_e \varphi_{x_h}(x_e)\frac{\partial \varphi_{x_h}(x_e)}{\partial x_h}\nonumber\\
=&\frac{\partial}{2\partial x_h}\int dx_e \Big(\varphi_{x_h}(x_e)\Big)^2,\\
J(x_h)=&\int dx_{e}\varphi_{x_h}(x_e)\frac{\partial^2\varphi_{x_h}(x_e)}{\partial x_h^2}\nonumber\\
=&\frac{\partial^2}{2\partial x_h^2}\int dx_e\Big(\varphi_{x_h}(x_e)\Big)^2-\int dx_e
\left( \frac{\partial \varphi}{\partial x_h} \right)^2.
\end{align}
Normalization condition ensures that $I(x_h)=0$ and that the first term in the right-hand side in
the expression for $J(x_h)$ is zero.

The fact that the equation for $\Phi(x_h)$ is closed is a consequence
of our choice  Eq. (\ref{form}) of the wave function $\Psi(x_e,x_h)$ in the form of the product.
By making this choice, we neglected the excited states of the electron wave function.
This choice is justified if the typical electron energy, $E_e(\kappa_h)$, is much
bigger than the size-quantization energy, $\sim \hbar^2/m_hL^2$,   of the hole. Upon setting $I(x_h)=0$ we cast Eq. (\ref{effective1}) into the form of the
Schr{\"o}dinger equation
\begin{equation}
\label{effective1}
-\frac{\hbar^2}{2m_h}\frac{\partial^2\Phi }{\partial x_h^2}+ V(x_{h})\Phi =E\Phi,
\end{equation}
with effective potential, $V(x_{h})$, representing the sum
\begin{equation}
\label{effective2}
V(x_{h})=\frac{\hbar^2}{2m_h}\int dx_e
\left( \frac{\partial \varphi_{x_h}(x_e)}{\partial x_h} \right)^2-\frac{\hbar^2}{2m_e}\Big(k_{h}(x_h)\Big)^2.
\end{equation}


\section{Comparing contributions to the effective potential}

At distances much bigger than $a_B$ from the boundary the second term in $V(x_h)$ dominates, we thus have $\kappa_h=\frac{1}{a_B}$,
so that $V(x_h)=-E_B$. Upon approach to the right boundary $x_h=\frac{L}{2}$ the term $\frac{\hbar^2\kappa_h^2}{2m_e}$ falls off and finally vanishes at $x_h= \frac{L-a_B}{2}$.
Near this point the first term in $V(x_h)$ dominates  the effective potential. To
trace the crossover between the first and second terms, it is sufficient to use the asymptotic expressions Eqs. (\ref{linearly}) and (\ref{normalized}).
Differentiating
Eq. (\ref{normalized}) with respect to $x_e$ we get
\begin{equation}
\label{derivative}
\frac{\partial \varphi_{x_h}(x_e)}{\partial x_h}=\bigl[1+2\kappa_h(x_e-x_h)\bigr]
\exp\bigl[\kappa_h(x_e-x_h)   \bigr]\frac{d\kappa_h}{dx_h}.
\end{equation}
Expressing $\frac{d\kappa_h}{dx_h}$ from Eq. (\ref{linearly}) we have

\begin{align}
\label{square}
&\Biggl( \frac{\partial \varphi_{x_h}(x_e)}{\partial x_h}  \Biggr)^2\nonumber\\
&=\frac{8}{\kappa_ha_B^4}\Bigl[1+2\kappa_h(x_e-x_h)\Bigr]^2
\exp\bigl[2\kappa_h(x_e-x_h)   \bigr].
\end{align}
Then the integration over $x_e$ yields

\begin{align}
\label{1holepotential}
&\frac{\hbar^2}{2m_h}\int dx_e
\left( \frac{\partial \varphi_{x_h}(x_e)}{\partial x_h} \right)^2\nonumber\\
&=\frac{2\hbar^2}{m_h\kappa_h^2a_B^4}
=\frac{\hbar^2}{8m_h\big[L-a_B-2x_h     \big]^2}.
\end{align}
Thus, the first term in $V(x_h)$ falls off quadratically away from the threshold
$x_h=\frac{1}{2}\left(L-a_B   \right)$. It should be compared to
the second term, $\frac{\hbar^2\kappa_h^2}{2m_e}$. Using the threshold behavior
Eq. (\ref{linearly})   of $\kappa_h$, we have

\begin{equation}
\label{2holepotential}
\frac{\hbar^2\kappa_h^2}{2m_e}=\frac{2\hbar^2}{m_ea_B^4}\bigg[L-a_B-2x_h   \bigg]^2.
\end{equation}
Comparing Eqs. (\ref{1holepotential}) and (\ref{2holepotential})
we conclude that the crossover from the first to the second term
in $V(x_h)$ takes place at
\begin{equation}
\label{crossover}
L-a_B-2x_h=\frac{a_B}{2}\left(\frac{m_e}{m_h}\right)^{1/4}.
\end{equation}
We see that crossover takes place in the threshold region, i.e. at $\left( L-a_B-2x_h\right)\ll a_B$. This is ensured by the smallness of the ratio $m_e/m_h$. 
Note, that this smallness also justifies the above use
of the asymptotic expressions for $\varphi_h(x_e)$
and for $\kappa_h(x_h)$.

\section{Energy levels}
Semiclassical quantization condition for the particle with mass, $m_h$, moving in the potential, $V(x_h)$, see Fig.~\ref{figure1}, reads

\begin{align}
\label{Bohr-Sommerfeld}
&2\left(\frac{2m_h}{\hbar^2}\right)^{1/2}\int\limits_0^{x_t}
dx_h\Bigg[E_N +\frac{\hbar^2\bigl(\kappa_h(x_h)\bigr)^2}{2m_e}   \Bigg]^{1/2}\nonumber\\
&=\pi\left(N+\frac{1}{2}   \right).
\end{align}
where $x_t$ is the turning point defined as
$\hbar\kappa_h(x_t)=\left(2m_e|E_N|    \right)^{1/2}$.
Using  Eq.  (\ref{limit})  we get the following expression for $x_t$
\begin{equation}
\label{x_t}
x_t=\frac{L}{2} +\frac{a_B}{2}\Biggl(\frac{E_B}{\vert E_N\vert}\Biggr)^{1/2}\ln\Biggl[1-\biggl(\frac{\vert E_N\vert}{E_B}  \biggr)^{1/2}\Biggr].
\end{equation}
Since $\kappa_h$ changes only in the vicinity of the upper limit and is
equal to $\frac{1}{a_B}$ otherwise, it is convenient to isolate the
constant part
\begin{align}
\label{isolate}
\Bigg[E_N +\frac{\hbar^2\kappa_h^2}{2m_e}   \Bigg]^{1/2}=\Bigg[E_N +\frac{\hbar^2}{2m_ea_B^2}   \Bigg]^{1/2}\nonumber\\
+\frac{\frac{\hbar^2}{2m_e}\bigl(\kappa_h^2-\frac{1}{a_B^2}\bigr)}
{\biggl(E_N+\frac{\hbar^2\kappa_h^2}{2m_e}\biggr)^{1/2}+
\biggl(E_N+\frac{\hbar^2}{2m_ea_B^2}\biggr)^{1/2}}.
\end{align}
Integration of the first term is elementary. To perform integration
of the second term it is convenient to switch from the variable $x_h$ to
 the dimensionless variable $u=\kappa_ha$. To do so, we differentiate both sides of Eq.~(\ref{limit}).
This yields

\begin{equation}
\label{differential}
\frac{du}{2u^2}\bigg[\ln\left(1-u\right)+\frac{u}{1-u}\bigg]=-\frac{dx_h}{a_B}.
\end{equation}
Then the condition Eq. (\ref{Bohr-Sommerfeld}) takes the form
\begin{align}
\label{Bohr-Sommerfeld2}
&\frac{\pi}{2}\left(N+\frac{1}{2}\right)
\Biggl(\frac{m_e}{m_h}\Biggr)^{1/2}
 \nonumber\\
&=\left(1-q_N\right)^{1/2}\Bigg[\frac{L}{2a_B}+\frac{\ln\left(1-q_N^{1/2}   \right)}{2q_N^{1/2}}\Bigg]-I(q_N).
\end{align}
where the  function $I(q_N)$ is defined as
\begin{align}
\label{integral}
&I(q_N)\nonumber =\\
&\int\limits_{q_N^{1/2}}^1\frac{du}{2u^2}\bigg[\ln\left(1-u\right)+\frac{u}{1-u}\bigg]
\frac{1-u^2}{\left(u^2-q_N\right)^{1/2}+\left(1-q_N\right)^{1/2}}.
\end{align}
The parameter
\begin{equation}
\label{notation}
q_N=\frac{\vert E_N\vert}{E_B}
\end{equation}
is the dimensionless exciton energy, so that $1-q_N$ has a meaning of the
dimensionless size-quantization energy. In the limit of large $L$ the size-quantization energy is much smaller than $E_B$, so that $\left(1-q_N\right)\ll 1$. This, in turn, means that the lower limit in the integral Eq. (\ref{integral})
is close to $1$. Then the integral $I(q_N)$ can be evaluated asymptotically
in small parameter $1-q_N$, which yields
\begin{equation}
\label{I_N}
I(q_N)=\frac{\left(1-q_N\right)^{1/2}}{2}\ln2.
\end{equation}
Substituting this expression in the right-hand side of Eq.~(\ref{Bohr-Sommerfeld2}), we arrive at a closed equation for $q_N$, which can be also viewed as the equation for
the electron energies, $E_N$. Solving this equation in the limit of large $L$,
we get
\begin{equation}
\label{E_N}
E_N=-E_B +\frac{\hbar^2\pi^2n^2}{2m_h\bigl(L-l_N\bigr)^2},
\end{equation}
where $l_N$ is given by
\begin{align}
\label{l}
&l_N=a_B\ln\Biggl[\frac{E_B}{\frac{\pi^2\hbar^2}{2m_hL^2}\Bigl(N+\frac{1}{2}   \Bigr)^2}    \Biggr]\nonumber\\
&=2a_B\Biggl[\ln\frac{L}{\pi a_B  \Bigl(N+\frac{1}{2}   \Bigr)}
+\frac{1}{2}\ln\frac{m_h}{m_e}           \Biggr].
\end{align}
We see that the result Eq. (\ref{E_N}) essentially reproduces Eq.~(\ref{formula})
but with the width of the dead layer specified. We see that the width, $l_N$,
grows with $L$, which is somewhat non-trivial.

\section{Discussion}
It is seen from Eq. (\ref{l}) that the arguments of both logarithms in the right-hand side
are large. This suggests that the size of the dead layer exceeds the exciton radius {\em parametrically}.
On the other hand, it follows from Eq. (\ref{1holepotential}) that the
effective repulsive potential for a hole diverges at $x_h=\frac{1}{2}\left(L-a_B\right)$,
i.e. at a distance $\frac{a_B}{2}$ from the ``hard wall" at $x_h=\frac{L}{2}$.

The fact that the turning point, $x_t$, for the hole motion, given by Eq. (\ref{x_t}),
is parametrically further away from the boundary than $a_B$ suggests
the following picture of the exciton reflection.
As a hole slowly moves towards the boundary,
it is orbited by a fast-moving electron. The energy of the system is $-E_B$.
Upon the approach to the boundary, the electron  ``senses" the
boundary by virtue of the no-escape boundary condition. As a result,
the energy of the system electron+hole increases. This increase acts
as a barrier for the hole, from which the hole is reflected being accompanied by the electron.

Our conclusion that $l\gg a_B$ suggests that the above assumption about the
short-range character of the electron-hole attraction can be relaxed. Moreover,
we can extend the above consideration to the 3D case. In the Appendix we derive
the form of potential barrier, ${\tilde V}(x_h)$, which enters into the quantization
condition

\begin{align}
\label{Bohr-Sommerfeld3}
&2\left(\frac{2m_h}{\hbar^2}\right)^{1/2}\int\limits_0^{x_t}
dx_h\Bigg[E_N+E_B -{\tilde V}(x_h)\Bigg]^{1/2}\nonumber\\
&=\pi\left(N+\frac{1}{2}   \right).
\end{align}
With the help of Eq. (\ref{A5}) we find the positions
of the turning points

\begin{equation}
\label{turning}
x_t=\frac{L}{2}+\frac{a_B}{2}\ln\Biggl(1-\frac{|E_N|}{E_B}   \Biggr),
\end{equation}
which appears to be similar to the 1D expression  Eq.~(\ref{x_t}). Repeating the steps in the previous section,
it can be readily shown that our main result Eq. (\ref{l})   for the width of the dead layer remains valid in three
dimensions.
\appendix
\section{Effective potential in 3D}

In this Appendix we calculate the correction to the ground state energy
of electron in the field of a hole due to the presence of a boundary at $x=\frac{L}{2}$.
We start by writing down the Schr{\"o}dinger equations with and without the boundary
\begin{equation}
\label{A1}
-\frac{\hbar^2}{2m_e}\Delta \psi_0 -\frac{e^2}{\vert {\bf r}_e-{\bf r}_h  \vert}\psi_0=-E_B\psi_0,
\end{equation}
\begin{equation}
\label{A2}
-\frac{\hbar^2}{2m_e}\Delta \psi -\frac{e^2}{\vert {\bf r}_e-{\bf r}_h  \vert}\psi +W\big(x_e-\frac{L}{2}\big)\psi =E\psi,
\end{equation}
where the function $W\big(x_e-\frac{L}{2}\big)$ describes a barrier which ensures that
electron does not penetrate into the region $x_e> \frac{L}{2}$.

Multiplying Eq. (\ref{A1}) by $\psi({\bf r})$ and Eq. (\ref{A2}) by $\psi_0({\bf r})$,
subtracting the two, and integrating over the domain $x_e<\frac{L}{2}$,
we find
\begin{align}
\label{A3}
E(x_h)+E_B=&-\frac{\hbar^2}{2m_e}\int\limits_{x_e=\frac{L}{2}}dy_e dz_e
\Bigg(\psi \frac{\partial \psi_0}{\partial x_e}-\psi_0\frac{\partial \psi}{\partial x_e}    \Bigg)\nonumber\\
&\approx \frac{\hbar^2}{4m_e}
\int\limits_{x_e=\frac{L}{2}}dy_e dz_e \frac{\partial \psi_0^2}{\partial x_e}.
\end{align}
In Eq. (\ref{A3}) we took into account that the first term in the brackets is zero,
since $\psi({\bf r})=0$ at the boundary. We have also set  $\psi({\bf r})=\psi_0({\bf r})$
in the second term.
Integration over the plane $x_e=\frac{L}{2}$ in the right-hand side yields the correction to the binding energy of the exciton due to the presence of the boundary.

The ground state wave function of an electron in the field of a hole which is located  at $(x_h,0,0)$
has the form
\begin{equation}
\label{A4}
\psi_0({\bf r}_e)=\frac{1}{\left( \pi a_B^3   \right)^{1/2}}\exp\Bigg\{-\frac{\Bigl[y_e^2+z_e^2+\left(x_e-x_h \right)^2  \Bigr]^{1/2}    }{a_B}    \Bigg\}.
\end{equation}
With $\psi_0({\bf r}_e)$ given by Eq. (\ref{A4}),
the integration in Eq.~(\ref{A3}) can be performed
explicitly in polar coordinates
\begin{align}
\label{A5}
&E(x_h)+E_B\nonumber\\
&=\frac{\hbar^2\big(\frac{L}{2}-x_h\big)}{2m_ea_B^4}
\int\limits_0^\infty d\rho^2\frac{\exp\Big[-\frac{2}{a_B}\Big(\rho^2+\big(\frac{L}{2}-x_h\big)^2\Big)^{1/2}      \Big]}
{\Big[\rho^2+\big(\frac{L}{2}-x_h\big)^2\Big]^{1/2}}\nonumber\\
&=\frac{\hbar^2\big(\frac{L}{2}-x_h\big)}{2m_ea_B^3}
\exp\Big[-\frac{2}{a_B}\Big(\frac{L}{2}-x_h\Big)\Big]={\tilde V}(x_h).
\end{align}
The result Eq. (\ref{A5}) defines the form of the barrier from which the
exciton is reflected.

\section{Acknowledgements}

\vspace{2mm}

The work was supported by the Department of Energy,
Office of Basic Energy Sciences, Grant No.  DE- FG02-
06ER46313.


\begin{thebibliography}{30}

\bibitem{pekar}
S. I. Pekar, ``The theory of electromagnetic waves in a crystal in which
excitons are produced,"
Zh. Eksp. Teor. Fiz. {\bf 33}, 1022 (1957) [Sov. Phys.
JETP {\bf 6}, 785 (1958)].

\bibitem{Hopfield} J. J. Hopfield and D. G. Thomas,
 ``Theoretical and Experimental Effects of Spatial Dispersion on the Optical Properties of Crystals,"
 Phys. Rev. {\bf 132}, 563 (1963).


\bibitem{1974}
F. Evangelisti, A. Frova, and F. Patella,
``Nature of the dead layer in CdS and its effect on exciton reflectance spectra,"
Phys. Rev. B{\bf 10},  4253 (1974).





\bibitem{Bassani}
A. Tredicucci, Y. Chen, F. Bassani, J. Massies, C. Deparis,
and G. Neu, ``Center-of-mass quantization of
excitons and polariton interference in GaAs thin layers,"
Phys. Rev. B {\bf 47} 10348, (1993).

\bibitem{DelSole2}
``Wannier-Mott excitons in semi-infinite crystals: Wave functions and normal-incidence reflectivity,"
Phys. Rev. B {\bf 25}, 3714 (1982).

\bibitem{DelSole1}
A. D'Andrea and R. Del Sole,
``New insight on exciton-polaritons based
on a microscopic approach,"  Phys. Rev. B {\bf 29}, 4782 (1984).
\bibitem{DelSole}
D. Viri, R. Del Sole, and A. D'Andrea,
``Exciton-free-layer depth as a function of the electron-hole mass ratio,"
Phys. Rev. B {\bf 48}, 9110 (1993).

\bibitem{Bastard} S. Jaziri, G. Bastard, and R. Bennaceur,
``Centre-of-mass quantization of excitons in
GaAs quantum boxes,"
Semicond. Sci. Technol. {\bf 8}, 670 (1993).


\bibitem{Cardona}
S. Tsoi, X. Lu, A. K. Ramdas, H. Alawadhi, M. Grimsditch, M. Cardona, and R. Lauck,
``Isotopic-mass dependence of the A, B, and C excitonic band gaps in
ZnO at low temperatures, Phys. Rev. B {\bf 74}, 165203 (2006).



\bibitem{Ivchenko} E. S. Khramtsov, P. S. Grigoryev, D. K. Loginov, I. V. Ignatiev,
Yu. P. Efimov, S. A. Eliseev, P. Yu. Shapochkin, E. L. Ivchenko, and M. Bayer,
``Exciton spectroscopy of optical reflection from wide quantum wells,"
Phys. Rev. B {\bf 99}, 035431 (2019).
\bibitem{Poddubny} A. N. Poddubny,
``Quasiflat band enabling subradiant two-photon bound states,"
Phys. Rev. A {\bf 101}, 043845 (2020).

\bibitem{ekimov}
A. I. Ekimov and  A. A. Onushchenko,
``Quantum size effect in three-dimensional
microscopic semiconductor crystals,"
JETP Lett. {\bf 34}, 345 (1981).

\bibitem{Brus} R. Rosetti, S. Nakahara, and L. E. Brus,
``Quantum size effects in the redox potentials,
resonance Raman spectra, and electronic spectra
of CdS crystallites in aqueous solution,"
J. Chem. Phys. {\bf 79}, 1986 (1983).




\bibitem{Fonoberov}
V. A. Fonoberov and A. A. Balandin,
``Radiative lifetime of excitons in ZnO nanocrystals:
The dead-layer effect," Phys. Rev. B {\bf 70}, 195410 (2004).


\bibitem{nano1}] L. Protesescu, S. Yakunin, M. I. Bodnarchuk, F. Krieg, R.
Caputo, C. H. Hendon, R. X. Yang, A. Walsh, and M. V.
Kovalenko,
``Nanocrystals of Cesium Lead Halide Perovskites (CsPbX3, X = Cl, Br, and I): Novel Optoelectronic Materials Showing Bright Emission with Wide Color Gamut,"
Nano Lett. {\bf 15}, 3692 (2015).

\bibitem{nano2}
Ch. Yin, L. Chen, N. Song, Y. Lv, F. Hu, Ch. Sun, W. W. Yu, Ch. Zhang, X. Wang, Y. Zhang, and M. Xiao,
``Bright-Exciton Fine-Structure Splittings in Single Perovskite Nanocrystals,"
Phys. Rev. Lett. {\bf 119}, 026401 (2017).

\bibitem{nano3}
    M. A. Becker, R. Vaxenburg, G. Nedelcu, P. C. Sercel, A. Shabaev,
    M. J. Mehl, J. G. Michopoulos, S. G. Lambrakos, N. Bernstein, J. L. Lyons, Th. St{\"o}ferle, R. F. Mahrt, M. V. Kovalenko, D. J. Norris, G. Rain{\`o}, and A. L. Efros,
    ``Bright triplet excitons in caesium lead halide perovskites,"
    Nature {\bf 553}, 189 (2018).
\bibitem{nano4}
Q. A. Akkerman, G. Rain{\`o}, M. V. Kovalenko, and L. Manna,
``Genesis, challenges and opportunities for colloidal
lead halide perovskite nanocrystals,"
Nature Mater. {\bf 7}, 394 (2018).

\bibitem{nano5}
M. O. Nestoklon, S. V. Goupalov, R. I. Dzhioev, O. S. Ken, V. L. Korenev, Yu. G. Kusrayev, V. F. Sapega, C. de Weerd, L. Gomez, T. Gregorkiewicz, J. Lin, K. Suenaga, Y. Fujiwara, L. B. Matyushkin, and I. N. Yassievich,
``Optical orientation and alignment of excitons in ensembles of inorganic perovskite nanocrystals,"
Phys. Rev. B {\bf 97}, 235304 (2018).


\end{thebibliography}
\end{document}